\documentclass[a4paper,10pt,twoside]{just}

\usepackage{multicol}
\usepackage{graphicx}
\usepackage{booktabs}
\usepackage{amssymb,bm,mathrsfs,bbm,amscd}
\usepackage[tbtags]{amsmath}
\usepackage{lastpage}
\usepackage{CJK}

\begin{document}
\begin{CJK*}{GBK}{song}

\fancyhead[c]{\small  10th International Workshop on $e^+e^-$ collisions 
from $\phi$ to $\psi$ (PhiPsi15)}
 \fancyfoot[C]{\small PhiPsi15-\thepage}


\title{A resolution of the puzzle of low $V_{us}$ values from
inclusive flavor-breaking sum rule analyses of hadronic $\tau$ decay 
data\thanks{Supported by the Natural Sciences and Engineering
Research Council of Canada and the Australian Research Council}}

\author{K. Maltman$^{1,2}$\email{kmaltman@yorku.ca}
\quad R.J. Hudspith$^1$\email{renwick.james.hudspith@gmail.com}
\quad R. Lewis$^1$\email{randy.lewis@yorku.ca}
\quad C.E. Wolfe$^1$\email{wolfe@yorku.ca}
\quad J. Zanotti$^2$\email{james.zanotti@adelaide.edu.au}}
\maketitle

\address{%
$^1$ York University, Toronto, ON, M3J 1P3, Canada\\
$^2$ CSSM, University of Adelaide, Adelaide, SA, 5005, Australia\\}

\begin{abstract}
Continuum and lattice methods are used to investigate systematic issues 
in the sum rule determination of $V_{us}$ using inclusive hadronic $\tau$ 
decay data. Results for $V_{us}$ employing assumptions for $D>4$ OPE 
contributions used in previous conventional implementations of this 
approach are shown to display unphysical dependence on the sum rule
weight, $w$, and choice of upper limit, $s_0$, of the relevant experimental
spectral integrals. Continuum and lattice results suggest a new 
implementation of the sum rule approach with not just $\vert V_{us}\vert$, 
but also $D>4$ effective condensates, fit to data. Lattice results 
are also shown to provide a quantitative assessment of truncation 
uncertainties for the slowly converging $D=2$ OPE series. The new sum
rule implementation yields $\vert V_{us}\vert$ results free of unphysical 
$s_0$- and $w$-dependences and $\sim 0.0020$ higher than that obtained
using the conventional implementation. With preliminary new experimental 
results for the $K\pi$ branching fraction, the resulting $\vert V_{us}\vert$ 
is in excellent agreement with that based on $K_{\ell 3}$, and
compatible within errors with expectations from three-family unitarity.
\end{abstract}

\begin{keyword}
$V_{us}$, lattice, sum rules
\end{keyword}

\begin{pacs}
12.15.Hh, 11.55.Hx, 12.38.Gc
\end{pacs}

\begin{multicols}{2}

\section{Introduction}
The conventional $\tau$ decay determination of $\vert V_{us}\vert$ 
employs finite-energy sum rules (FESRs) and flavor-breaking (FB)
combinations of inclusive hadronic $\tau$ decay data~\cite{gamizetal}.
With $\Pi^{(J)}_{V/A;ij}(s)$ the $J=0,1$ components of flavor
$ij=ud,us$, vector (V) or axial vector (A) current 2-point
functions, $\rho^{(J)}_{V/A;ij}(s)$ the corresponding spectral 
functions, and $\Delta\Pi_\tau \, \equiv\,
\left[ \Pi_{V+A;ud}^{(0+1)}\, -\, \Pi_{V+A;us}^{(0+1)}\right]$,
the FESR relation
\begin{equation}
\int_0^{s_0}w(s) \Delta\rho_\tau (s)\, ds\, =\,
-{\frac{1}{2\pi i}}\oint_{\vert
s\vert =s_0}w(s) \Delta\Pi_\tau (s)\, ds\ ,
\label{basicfesr}
\end{equation}
is valid for any $s_0$ and any analytic $w(s)$. The spectral
function of $\Delta\Pi_\tau$, $\Delta\rho_\tau$, is experimentally 
accessible in terms of the differential distribution, 
$dR_{V/A;ij}/ds$, of 
$R_{V/A;ij}\, \equiv\, \Gamma [\tau^- \rightarrow \nu_\tau
\, {\rm hadrons}_{V/A;ij}\, (\gamma )]/ \Gamma [\tau^- \rightarrow
\nu_\tau e^- {\bar \nu}_e (\gamma)]$. Explicitly~\cite{tsai71}
\begin{eqnarray}
&&{\frac{dR_{V/A;ij}}{ds}}\, =\, c^{EW}_\tau \vert V_{ij}\vert^2
\left[ w_\tau (s ) \rho_{V/A;ij}^{(0+1)}(y_\tau )\right.\nonumber\\
&&\left. \qquad\qquad\quad - w_L (y_\tau )\rho_{V/A;ij}^{(0)}(s) \right]
\label{basictaudecay}\end{eqnarray}
with $y_\tau =s/m_\tau^2$, $w_\tau (y)=(1-y)^2(1+2y)$,
$w_{L}(y)=2y(1-y)^2$, $c^{EW}_\tau$ a known constant,
and $V_{ij}$ the flavor $ij$ CKM matrix element. 
The RHS of Eq.~(\ref{basicfesr}) is treated using the OPE.

One uses the $J=0+1$ FESR Eq.~(\ref{basicfesr}), rather than the analogue 
involving the spectral function combination in Eq.~(\ref{basictaudecay}), 
because of the very bad behavior of the integrated $J=0$, $D=2$ OPE 
series~\cite{longprob}. $\rho_{V/A;ud,us}^{(0+1)}(s)$ is obtained after 
subtracting phenomenologically determined $J=0$ contributions from
$dR_{V/A;ud;us}/ds$. This subtraction is dominated by the
accurately known, non-chirally-suppressed $\pi$ and $K$ 
pole terms. Continuum $\rho_{V/A;ud}^{(0)}$ contributions are
$\propto (m_d\mp m_u)^2$ and numerically negligible. Small, 
but not totally negligible, $(m_s\mp m_u)^2$-suppressed 
continuum $\rho_{V/A;us}^{(0)}$ contributions are determined
using highly constrained dispersive and sum rule methods~\cite{jop,mksps}. 
With $\vert V_{ud}\vert$ fixed~\cite{ht14}, $\Delta\rho_\tau (s)$ is 
determined by experimental data and $\vert V_{us}\vert$. 
$\vert V_{us}\vert$ is then obtained using the OPE 
on the RHS and data on the LHS of Eq.~(\ref{basicfesr}).

From the distribution $dR^{(0+1)}_{V+A;ud,us}/ds$, obtained
after subtracting $J=0$ contributions, re-weighted $J=0+1$ versions,
$R^w_{V+A;ij}(s_0)\equiv \int_0^{s_0}ds\, {\frac{w(s)}
{w_\tau (s)}}\, {\frac{dR^{(0+1)}_{V+A;ij}(s)}{ds}}$, of $R_{V+A;ud,us}$,
may be constructed for any $w$ and $s_0\le m_\tau^2$. 
With $\delta R^{w,OPE}_{V+A}(s_0)$ the OPE representation of
$\delta R^w_{V+A}(s_0)\, \equiv\,
{\frac{R^w_{V+A;ud}(s_0)}{\vert V_{ud}\vert^2}}
\, -\, {\frac{R^w_{V+A;us}(s_0)}{\vert V_{us}\vert^2}}$,
one then has
\begin{equation}
\vert V_{us}\vert \, =\, \sqrt{R^w_{V+A;us}(s_0)/\left[
{\frac{R^w_{V+A;ud}(s_0)}{\vert V_{ud}\vert^2}}
\, -\, \delta R^{w,OPE}_{V+A}(s_0)\right]}\ .
\label{tauvussolution}\end{equation}
The resulting $\vert V_{us}\vert$ should be independent of $s_0$
and the choice of weight, $w$, provided all experimental data, 
and any assumptions employed 
in evaluating $\delta R^{w,OPE}_{V+A}(s_0)$, are reliable. Since 
integrated $D=2k+2$ OPE contributions scale as $1/s_0^k$,
problems with assumptions about higher $D$ non-perturbative 
contributions, e.g., will produce an unphysical $s_0$-dependence
in $\vert V_{us}\vert$.

The conventional implementation of Eq.~(\ref{tauvussolution})~\cite{gamizetal}
employs $w=w_\tau$ and $s_0=m_\tau^2$. With this choice, the spectral 
integrals $R^{w_\tau}_{V+A;ud,us}(m_\tau^2)$ are determinable from 
inclusive non-strange and strange hadronic $\tau$ branching fractions, 
but assumptions about higher dimension $D=6,8$ OPE contributions, in 
priniciple present for a degree $3$ weight like $w_\tau$, are unavoidable. 
Using a single $w$ and single $s_0$ precludes subjecting these assumptions to 
$w$- and $s_0$-independence tests. It is a long-standing puzzle that this 
implementation produces inclusive $\tau$ $\vert V_{us}\vert$ determinations 
$>3\sigma$ below 3-family-unitarity expectations (the most recent version, 
$\vert V_{us}\vert = 0.2176(21)$~\cite{tauvusckm14}, e.g., lies $3.6\sigma$
below the current unitarity expectation, 
$\vert V_{us}\vert =0.2258(9)$~\cite{ht14}).  
Tests of the conventional implementation, however, 
show sizeable $s_0$- and $w$-dependence~\cite{kmcwvus} (see also, e.g.,
the left panel, and solid lines in the right panel, of Fig.~\ref{fig1}),
indicating the existence of systematic problems in
the conventional implementation. The dashed lines in
the right panel show the results of the alternate implementation
discussed below.

\end{multicols}
\vskip .05in

\ruleup
\begin{center}
\includegraphics[width=.33\textwidth,angle=270]
{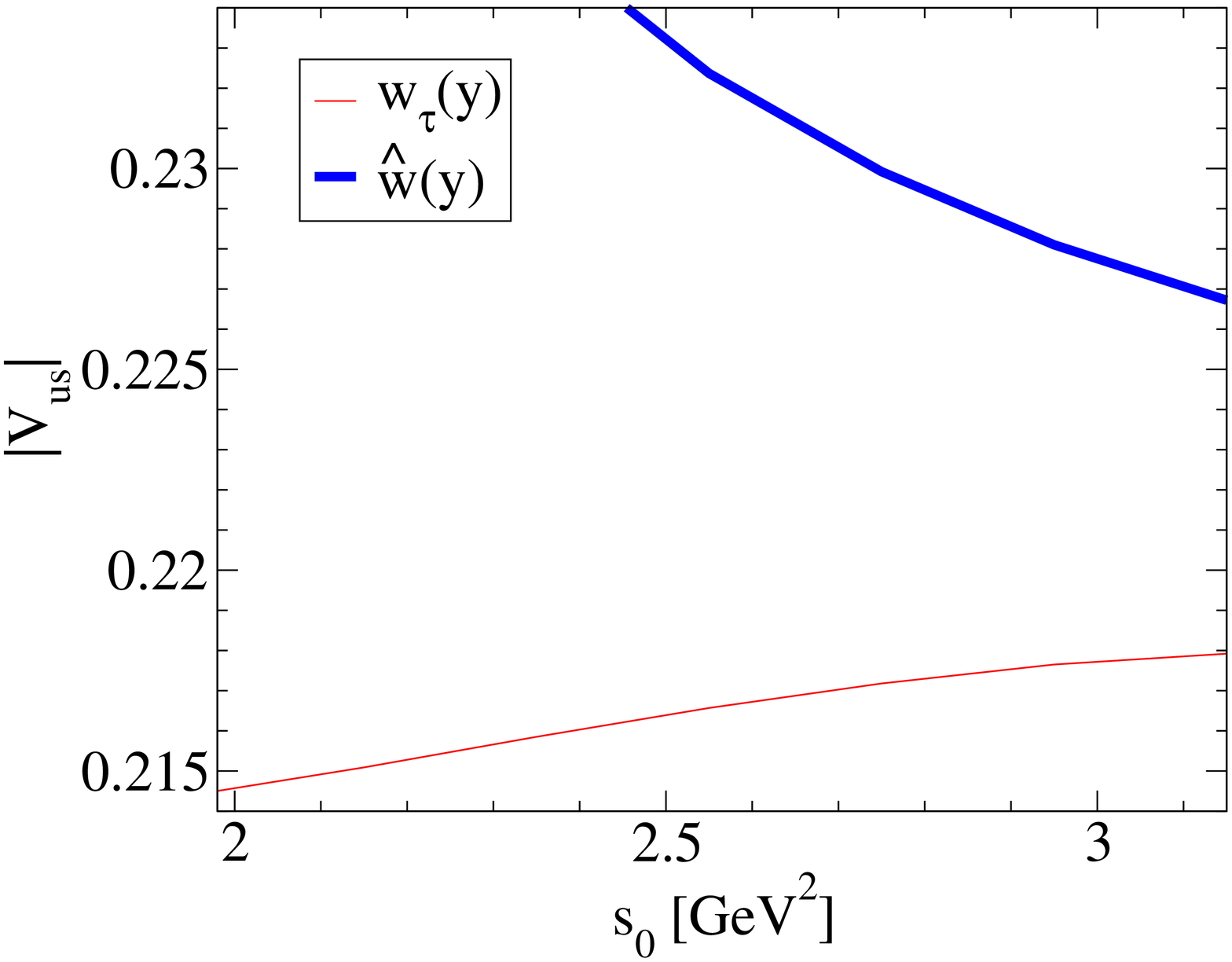}
\
\includegraphics[width=.33\textwidth,angle=270]
{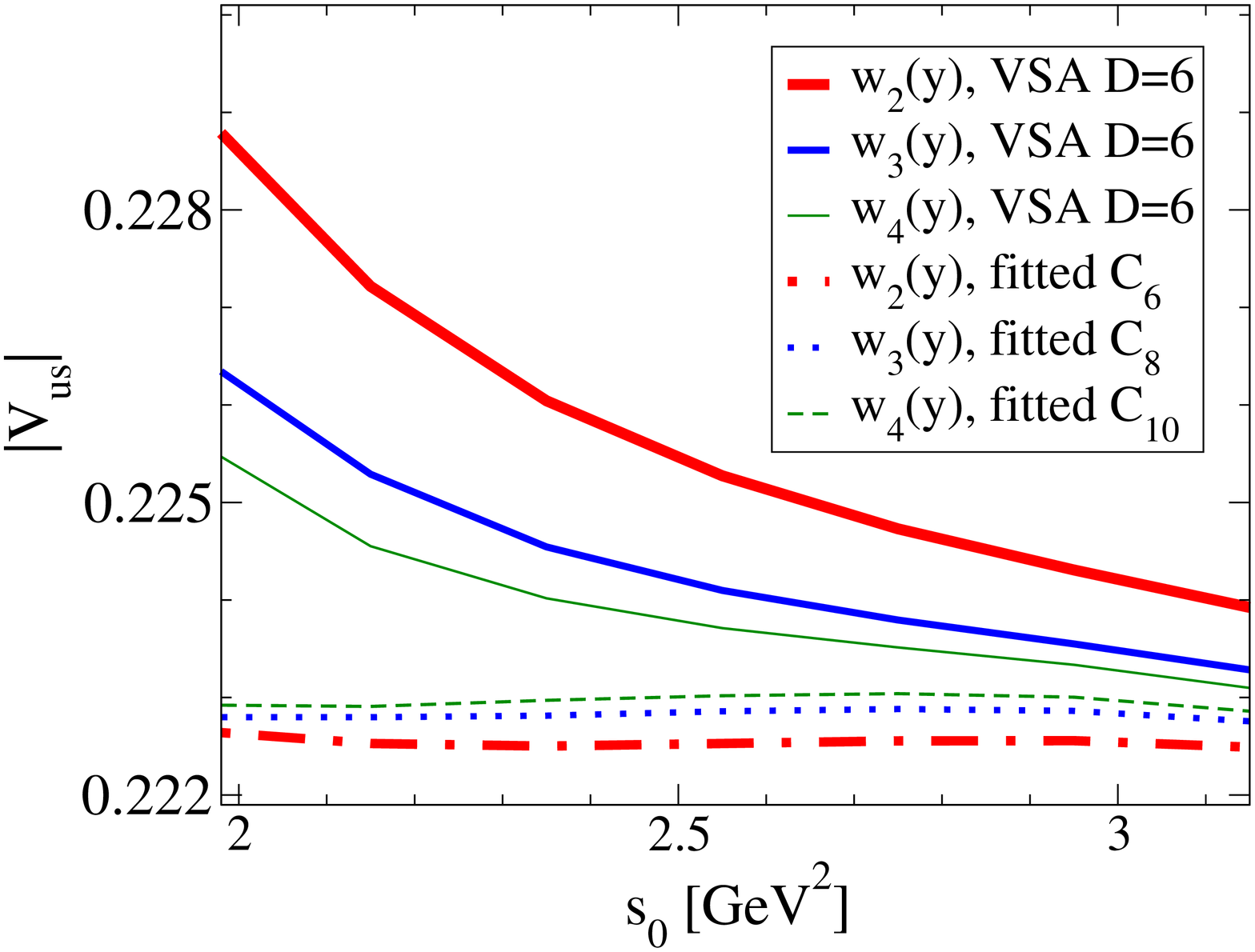}
\figcaption{\label{fig1} Left panel: $\vert V_{us}\vert$ from the
$w_\tau$ and $\hat{w}$ FESRs with standard~\cite{gamizetal}
OPE treatment (including CIPT for the $D=2$ series). Right panel: Comparison 
of conventional implementation results with those obtained
using central fitted $C_{6,8,10}$ values and the FOPT $D=2$ prescription 
favored by lattice results, for the weights $w_{2,3,4}$ defined in the
text.}
\end{center}
\ruledown
\vspace{0.5cm}
\begin{multicols}{2}

Two obvious theoretical systematic issues exist which might account for 
the observed $w$- and $s_0$-instabilities. The first concerns the treatment 
of $D=6$, $8$ OPE contributions. Both the conventional implementation and 
generalized versions just mentioned~\cite{kmcwvus}, estimate $D=6$ 
contributions using the vacuum saturation approximation (VSA) and neglect 
$D=8$ contributions. The VSA $D=6$ estimate is very small due to significant 
cancellations, both in the individual $ud$ and $us$ V+A sums and in the 
subsequent FB difference of these sums. With sizeable channel-dependent 
VSA breaking observed in the flavor $ud$ V and A channels~\cite{dv7}, 
such strong cancellations make the VSA estimate potentially quite 
unreliable. The second possibility concerns the slow convergence of the 
$D=2$ OPE series for $\Delta\Pi_\tau$. With $\bar{a}=\alpha_s(Q^2)/\pi$,
and $m_s(Q^2)$, $\alpha_s(Q^2)$ the running strange quark mass and 
coupling in the $\overline{MS}$ scheme, one has, to four loops~\cite{bckd2ope}
(neglecting $O(m^2_{u,d}/m^2_s)$ corrections)
\begin{eqnarray}
&&\left[\Delta\Pi_\tau (Q^2)\right]^{OPE}_{D=2}\, =\, {\frac{3}{2\pi^2}}\,
{\frac{m_s(Q^2)}{Q^2}} \left[ 1 + {\frac{7}{3}} \bar{a}
+ 19.93 \bar{a}\,^2\right.\nonumber\\
&&\left. \qquad\qquad\qquad\qquad \, + 208.75 \bar{a}\,^3 +\cdots\ \right]\ .
\label{d2form}\end{eqnarray}
Since $\bar{a}(m_\tau^2)\simeq 0.1$, convergence at the spacelike point on
$\vert s\vert = s_0$ is marginal at best, raising questions concerning the
choice of truncation order and truncation error estimates for the 
corresponding integrated series. 
The $D=2$ convergence/truncation issue is also evident in the significant 
difference (increasing from $\sim 0.0010$ to $\sim 0.0020$ between 3- 
and 5-loop truncation order) in $\vert V_{us}\vert$ results obtained using 
alternate (fixed-order (FOPT) and contour-improved (CIPT)) prescriptions
(prescriptions differing only by terms beyond the common truncation order) 
for the truncated integrated $D=2$ series~\cite{kmcwvus}.

In what follows, we first investigate the treatment of the $D=2$ OPE
series using lattice data for $\Delta\Pi_\tau$, then test the
$D=6,\, 8$ assumptions of the conventional implementation by comparing
FESR results for a judiciously chosen pair of weights, $w_\tau (y)$ and
$\hat{w}(y)=(1-y)^3$, $y=s/s_0$. Results obtained employing an alternate 
implementation of the FB FESR approach suggested by these investigations
are then presented.

\section{Lattice and continuum investigations of the OPE representation of 
$\Delta\Pi_\tau$}

Data for $\Delta\Pi_\tau (Q^2)$ can be generated over a wide range of 
Euclidean $Q^2$ using the lattice. The (tight) cylinder cut which
must be applied to avoid lattice artifacts at higher $Q^2$ has been 
determined, for the ensemble employed here, in a recent analysis 
aimed at using lattice current-current two-point function data to
determine $\alpha_s$~\cite{hlms15}. We first consider data at high 
enough $Q^2$ that $\left[\Delta\Pi_\tau\right]_{OPE}$ will be safely 
dominated by the leading $D=2$ and $4$ contributions. The latter are 
determined by light and strange quark masses and condensates and hence 
known. We use FLAG results for physical quark masses~\cite{FLAG} and 
GMOR for the light condensate. $\langle \bar{s}s\rangle$ then follows 
from $\langle \bar{s}s\rangle /\langle \bar{u}u\rangle$ (the HPQCD 
physical-$m_q$ version of this ratio~\cite{hpqcdcondratio} is easily 
translated to the $m_q$ of the ensemble employed using NLO 
ChPT~\cite{gl85}). We then consider various combinations of truncation 
order and log-resummation schemes for the $D=2$ OPE series, 
investigating whether a choice exists which produces a good match between 
the resulting $D=2+4$ OPE sum and the lattice data in the high-$Q^2$ region.

For this high-$Q^2$ study, we employ the RBC/UKQCD $n_f=2+1$, 
$32^3\times 64$, $1/a=2.38$ GeV, $m_\pi\sim 300$ MeV domain wall 
fermion ensemble~\cite{rbcukqcdfine11}. We find that 3-loop $D=2$ 
truncation with fixed-scale choice (the analogue of the FOPT FESR 
prescription) provides an excellent OPE-lattice match over a wide 
range of $Q^2$, extending from near $\sim 10$ GeV$^2$ down to just 
above $\sim 4$ GeV$^2$. The comparison is shown, for fixed scale 
choice $\mu^2=4\ GeV^2$, in the left panel of Fig.~\ref{fig2}, 
where, for ease of display, we plot results for the product 
$Q^2\, \Delta\Pi_\tau (Q^2)$ (this removes a factor of $1/Q^2$ 
present in the $D=2$ OPE contributions). The right panel of 
Fig.~\ref{fig2} shows the analogous comparison for 
the alternate local-scale ($\mu^2=Q^2$) choice (analogous to the 
CIPT FESR prescription). It is clear that the $Q^2$ dependence 
of the lattice data prefers the fixed-scale treatment of the $D=2$ series.

\end{multicols}
\vskip .05in
\ruleup
\begin{center}
\includegraphics[width=.33\textwidth,angle=270]
{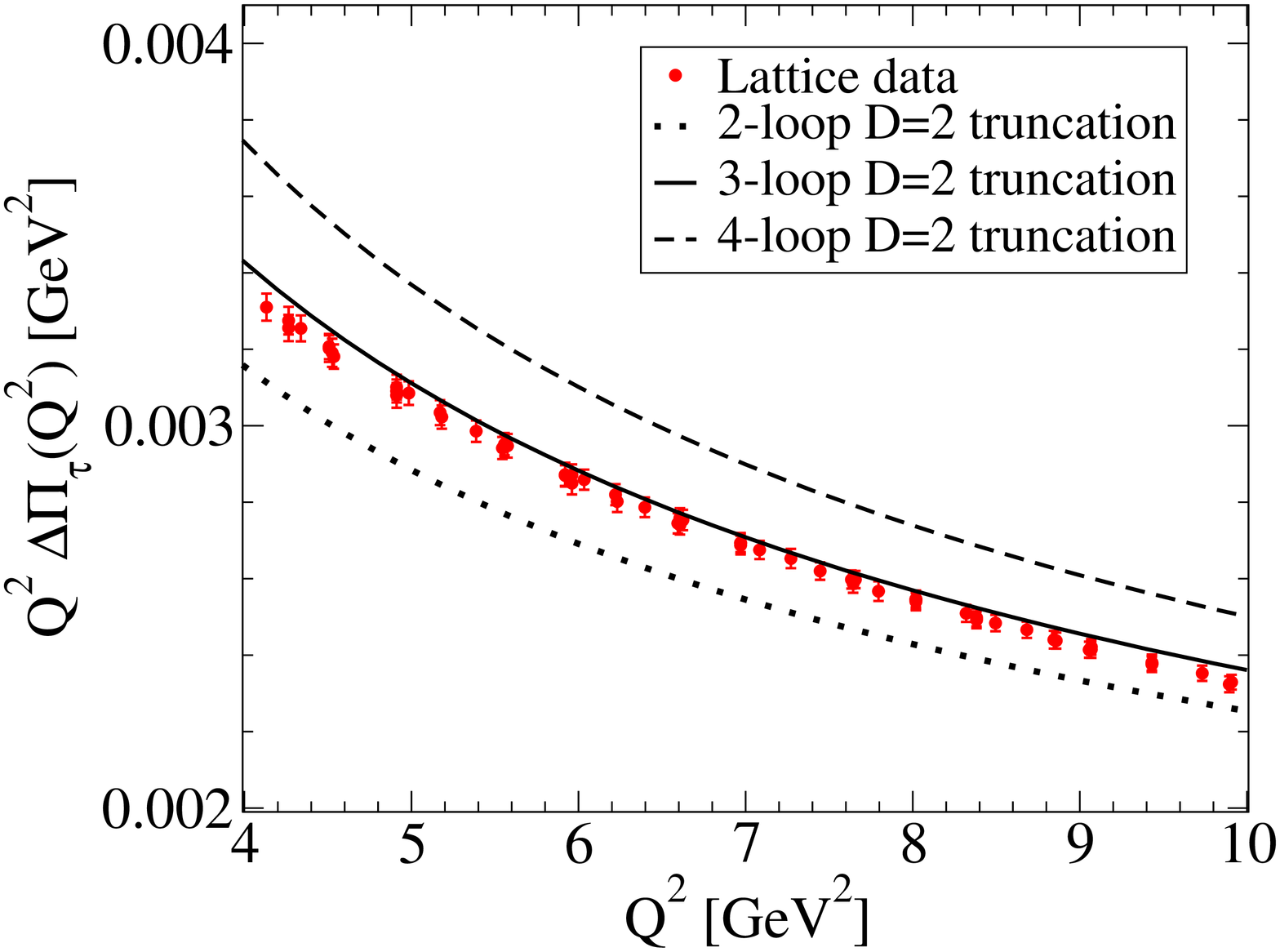}
\
\includegraphics[width=.33\textwidth,angle=270]
{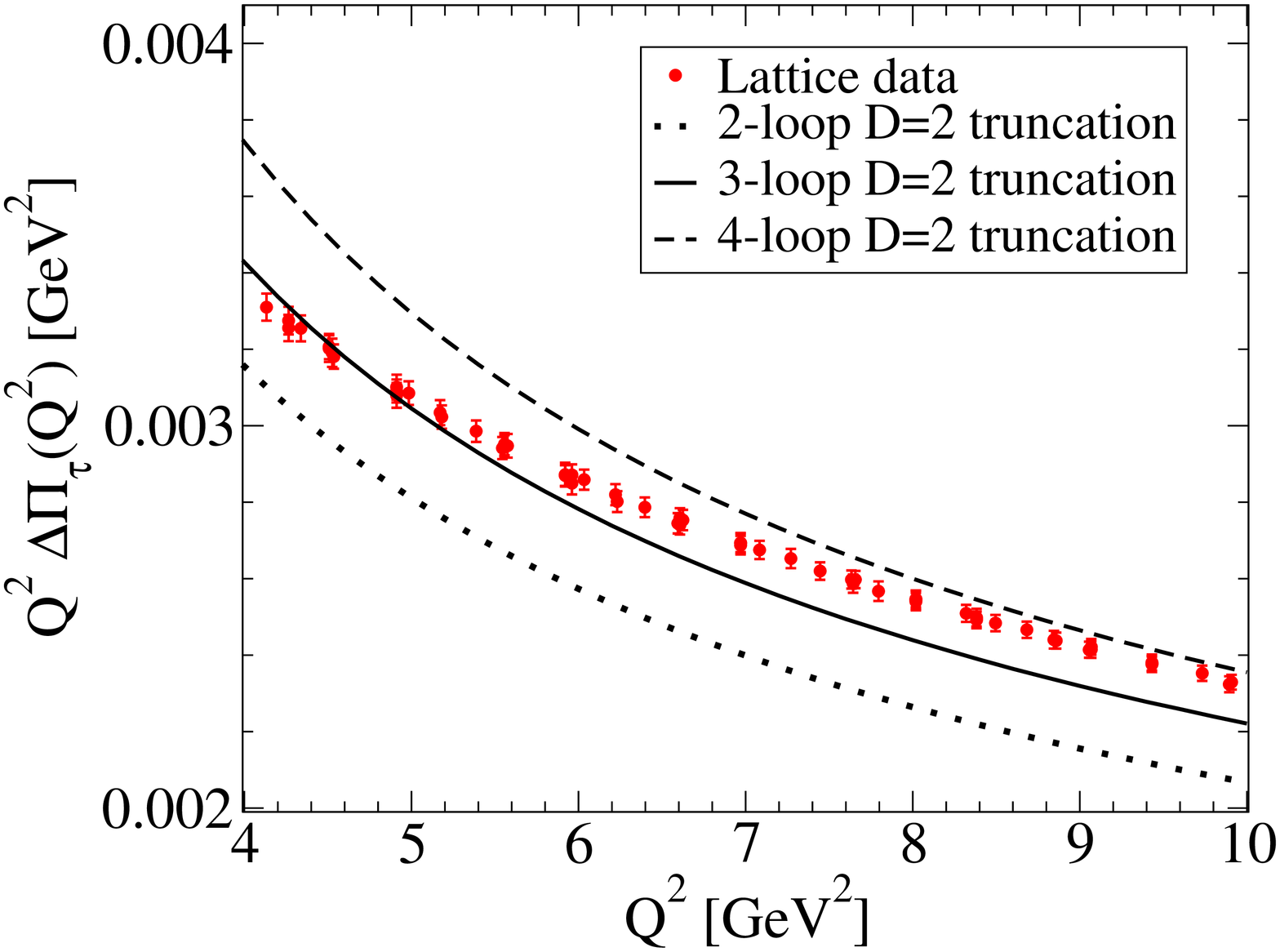}
\figcaption{\label{fig2}Comparison of lattice data and OPE 
$D=2+4$ expectations for $Q^2\, \Delta \Pi_\tau (Q^2)$,
for various truncation orders and either the fixed-scale treatment 
(left panel) or local-scale treatment (right panel) of the $D=2$ series.}
\end{center}
\ruledown
\vspace{0.5cm}
\begin{multicols}{2}

The lattice data also provides us with the possibility of investigating
the reliability of conventional methods for estimating the
theoretical error to be associated with the truncated OPE. This is of 
particular relevance given the very slow convergence of the $D=2$ series, 
which might raise doubts about the suitability of such conventional
estimates in the case of the $D=2$ truncation uncertainty.
Fig.~\ref{fig3} shows the $D=2+4$ OPE error band obtained using the 
3-loop-truncated, fixed-scale $D=2$ OPE treatment and such conventional 
OPE error estimates, taking into account uncertainties in
the input OPE parameters and using the magnitude of the last term
kept to estimate the $D=2$ series truncation uncertainty. One sees 
that, despite the very slow convergence of the $D=2$ series, the 
resulting conventionally determined error turns out to be
extremely conservative. 
\begin{center}
\includegraphics[width=.33\textwidth,angle=270]
{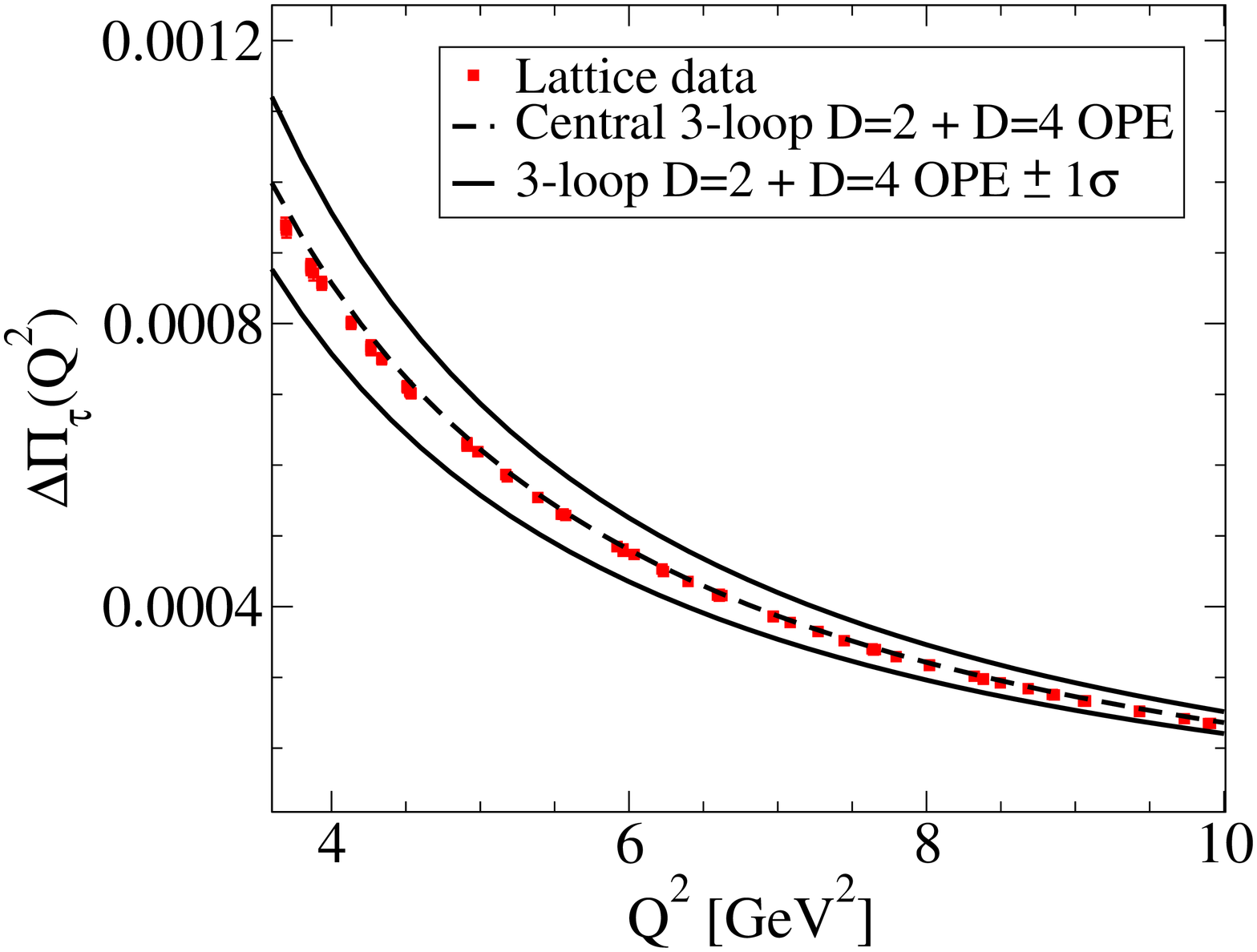}
\figcaption{\label{fig3}Lattice data and the $D=2+4$
OPE sum, with conventional OPE error estimates, for the 3-loop-truncated,
fixed-scale $D=2$ treatment}
\end{center}

We now switch our attention to lower $Q^2$, the goal being to 
use lattice data to test the assumptions about $D>4$ contributions 
underlying the standard implementation, namely that $D>6$ contributions 
are safely negligible and $D=6$ contributions can be reasonably well 
approximated using the VSA. Fig.~\ref{fig4} shows the comparison
of lattice data for $\Delta\Pi_\tau (Q^2)$ in the region below
$\sim 4\ GeV^2$ with two versions of the truncated OPE.
The dashed line represents the 3-loop-truncated, fixed-scale $D=2+4$ OPE
sum discussed above, which provides an excellent match to the lattice 
data at higher scales, while the solid line shows the result of 
supplementing the $D=2+4$ sum with the VSA estimate for the $D=6$ 
contribution. The results show clear evidence for the onset of $D>4$ 
contributions below $Q^2\sim 4\ GeV^2$ significantly larger than 
those obtained using the VSA estimate for $D=6$ and neglecting 
$D>6$ contributions. The VSA $D=6$ estimate is not only far too small 
in magnitude to bring the low-$Q^2$ truncated OPE into agreement with 
the lattice data but, in fact, moves the truncated OPE sum slightly in the
wrong direction. Unfortunately, the Euclidean lattice data provides 
no means of selectively isolating contributions of different $D>4$
in this lower $Q^2$ region. 
Further investigation of the higher $D$ question thus requires 
continuum FESR methods.

Our continuum FESR studies employ the $D=2,\, 4$ OPE treatment favored 
by lattice data, detailed above. Spectral integral input is as follows:
$\pi_{\mu2}$, $K_{\mu 2}$ and Standard Model expectations for the 
$\pi$ and $K$ pole contributions, recent ALEPH data for the continuum 
$ud$ V+A distribution~\cite{aleph13}, BaBar~\cite{babarkmpi0} and 
Belle~\cite{bellekspi} results for the $K^-\pi^0$ and $\bar{K}^0\pi^-$
distributions, BaBar results~\cite{babarkpipiallchg} for the
$K^-\pi^+\pi^-$ distribution, Belle results~\cite{bellekspipi} for the 
$\bar{K}^0\pi^-\pi^0$ distribution and 1999 ALEPH 
results~\cite{alephus99} for the combined distribution of those strange 
modes not remeasured by the B-factory experiments.

\begin{center}
\includegraphics[width=.33\textwidth,angle=270]
{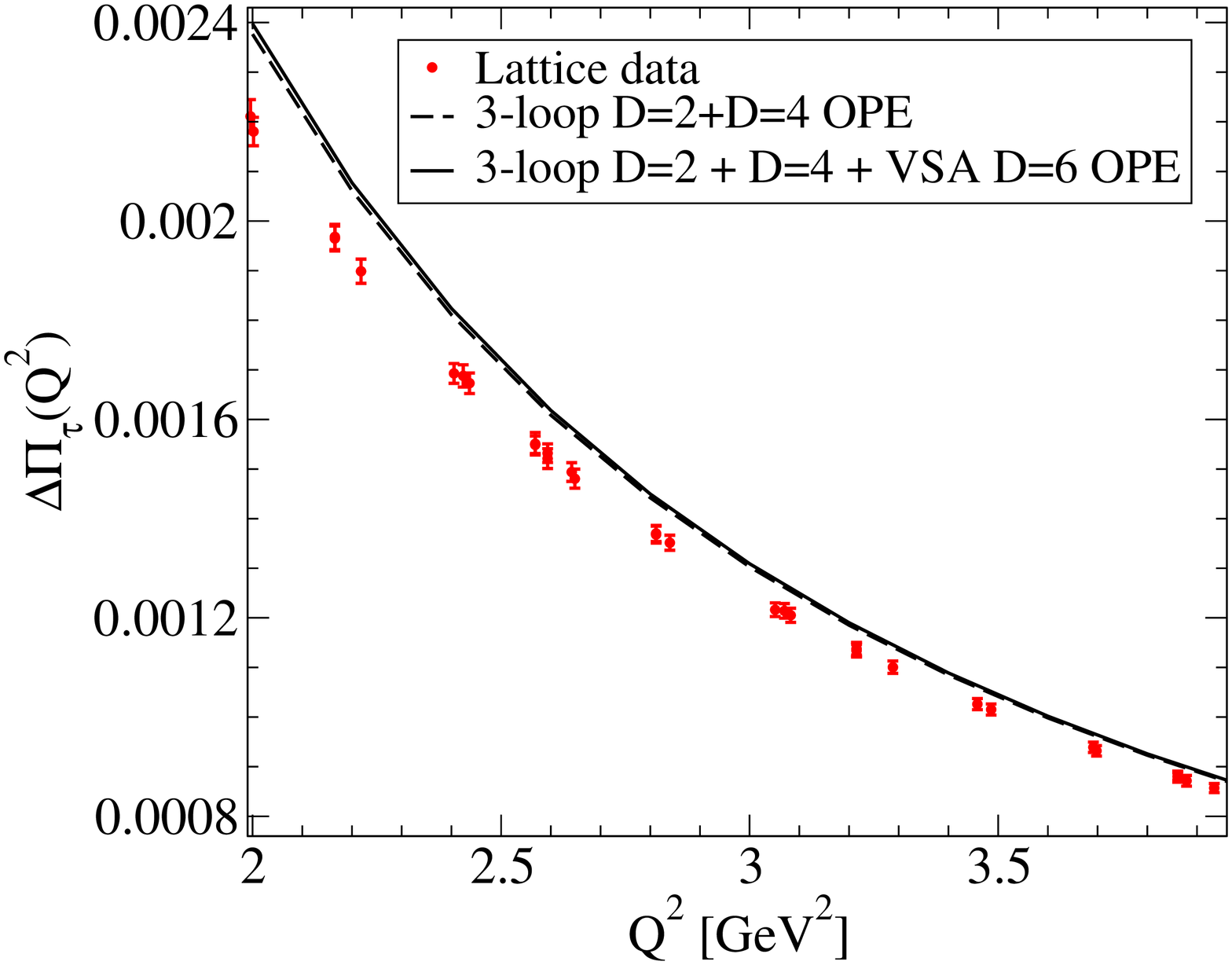}
\figcaption{\label{fig4}Comparison of lower-$Q^2$ lattice data with 
fixed-scale $D=2$ OPE based expectations, one employing just the
sum of $D=2$ and $4$ contributions, the other supplementing this
with the estimated $D=6$ contribution obtained using the VSA.}
\end{center}

The BaBar and Belle exclusive mode distributions are unit-normalized 
and must have their overall scales fixed using experimental branching 
fractions. In the 
results quoted below, HFAG strange exclusive mode branching fractions 
have been used, with the exception of the $K^-\pi^0$ mode, for which the 
updated version from the recent BaBar Adametz thesis~\cite{adametzthesis}
has been employed. A corresponding (very) small rescaling is applied to 
the continuum $ud$ V+A distribution to restore unitarity.

Neglecting $\alpha_s$-suppressed logarithmic corrections, $D>4$ OPE 
contributions to $\Delta\Pi_\tau (Q^2)$ can be written $\sum_{D> 4} C_D/Q^D$ 
with $C_D$ an effective dimension $D$ condensate. The degree 3 weights
$w_\tau (y)=1-3y^2+2y^3$ and $\hat{w}(y)=1-3y+3y^2-y^3$ generate integrated 
OPE contributions up to $D=8$ only. The integrated $D=6,\, 8$ results,
\begin{equation}
-{\frac{3C_6}{s_0^2}}\, -\, {\frac{2C_8}{s_0^3}}\ \ {\rm for\ }w_\tau 
\qquad {\rm and}\qquad 
{\frac{3C_6}{s_0^2}}\, +\, {\frac{C_8}{s_0^3}}\ \ {\rm for\ }\hat{w},
\label{wtauwhatcomp}\end{equation}
have $D=6$ contributions identical in magnitude but opposite in sign, 
and a $\hat{w}$ $D=8$ contribution similarly opposite in sign but half 
in magnitude that of $w_\tau$. Were the assumptions of the conventional 
implementation to be correct, with $D=6,\, 8$ contributions numerically 
negligible in the $w_\tau$ FESR, this will necessarily also be the case 
for the $\hat{w}$ FESR. The two FESRs should then produce results for
$\vert V_{us}\vert$ which not only agree, but are both $s_0$-independent. 
In contrast, if the $D=6$ and/or $8$ contributions to the $w_\tau$ FESR 
are not, in fact, negligible, the results for $\vert V_{us}\vert$ from the 
two FESRs, obtained assuming they are, should show $s_0$-instabilities of 
opposite sign, decreasing in magnitude with increasing $s_0$ for both.
The left panel of Fig.~\ref{fig1} shows the latter scenario to be the
one actually realized. The sizeable $s_0$- and weight-choice dependences 
demonstrate unambiguously the breakdown of the assumptions underlying 
the conventional implementation. The $3\sigma$ low $\vert V_{us}\vert$
results obtained employing them are thus afflicted with significant 
previously unquantified systematic uncertainties. 

\section{An alternate implementation of the FB FESR approach}
With previously employed approaches to estimating $D>4$ effective OPE 
condensates shown to be untenable, our only option is to fit these
condensates to data. This can be done only by exploiting the
fact that integrated OPE contributions of different $D$ scale
differently with $s_0$, and hence requires working with FESRs 
involving a range of values of the variable $s_0$. This precludes 
determining the required spectral integrals solely in terms of 
hadronic $\tau$ decay branching fractions. 

Considering FESRs based on different weights provides further tests of
possible theoretical systematics. To suppress duality violating 
contributions, we restrict our attention to weights having at least a double 
zero at $s=s_0$. The weights $w_N(y)=1-{\frac{N}{N-1}}y+{\frac{1}{N-1}}y^N$, 
$N\ge 2$~\cite{my08} are particularly convenient in this regard since 
the corresponding integrated OPE involves a single $D>4$ contribution 
(with $D=2N+2$). With $D=2+4$ OPE contributions under control (as discussed 
above) this leaves $\vert V_{us}\vert$ and the effective condensate
$C_{2N+2}$ as the only parameters to be determined. These are obtained
through a fit to the set of $w_N$-weighted spectral integrals in the 
chosen $s_0$ fit window. Further tests of the analysis are provided 
by verifying (i) that the $\vert V_{us}\vert$ obtained from the different 
$w_N$ FESRs are in good agreement and (ii) that the fitted $C_D$ 
are physically plausible, in the sense of showing FB cancellation 
relative to the results of Ref.~\cite{dv7} for the corresponding 
effective flavor $ud$ channel condensates. We have analyzed the 
$w_N$ FESRs for $N=2,\, 3,\, 4$ and verified that the results 
pass these self-consistency tests. 

In the right panel of Fig.~\ref{fig1} we display, as dashed/dotted 
lines, the results which follow from taking as input the central 
value for the condensate, $C_{2N+2}$, obtained from the $w_N$ FESR 
analysis, and solving Eq.~(\ref{tauvussolution}) for $\vert V_{us}\vert$
as a function of $s_0$. The results are displayed for each of the 
$w_2$, $w_3$ and $w_4$ FESR cases. The results make clear (i) the 
underlying excellent match between the fitted OPE and spectral 
integral sets, (ii) the excellent agreement between results 
for $\vert V_{us}\vert$ obtained from the different $w_N$ FESR
analyses, and (iii) the dramatic decrease in the $s_0$- and 
weight-dependence of the results for $\vert V_{us}\vert$ produced
by using $D>4$ OPE effective condensates fit to data in place of
those based on the assumptions of the conventional implementation.
One also sees that, as expected, the fitted $\vert V_{us}\vert$
lie between the $s_0$-unstable results produced by the conventional
implementation of the $w_\tau$ and $\hat{w}$ FESRs, and are
$\sim 0.0020$ higher than the results of the conventional $w_\tau$
implementation.

\end{multicols}
\vskip .1in
\begin{center}
\tabcaption{\label{tab1}Error budgets for the $w_2$, $w_3$ and
$w_4$ determinations of $\vert V_{us}\vert$, using the
3-loop-truncated, fixed-scale treatment of the $D=2$ OPE series}
\footnotesize
\begin{tabular*}{170mm}{@{\extracolsep{\fill}}lccc}
\toprule 
Error source&$\delta\vert V_{us}\vert$ ($w_2$ FESR)&
$\delta\vert V_{us}\vert$ ($w_3$ FESR)&
$\delta\vert V_{us}\vert$ ($w_4$ FESR)\\
\hline
$\delta\alpha_s$&0.00001&0.00004&0.00004\\
$\delta m_s(2\ GeV)$&0.00017&0.00019&0.00019\\
$\delta\langle m_s\bar{s}s\rangle$&0.00035&0.00035&0.00035\\
$\delta (long\ corr)$&0.00009&0.00009&0.00009\\
\hline
Experimental ($ud$)&0.00027&0.00028&0.00028\\
Experimental ($us$)&0.00226&0.00227&0.00227\\
\bottomrule
\end{tabular*}%
\end{center}
\vskip .1in
\begin{multicols}{2}

In Table~\ref{tab1}, we give the error budgets for the 
$\vert V_{us}\vert$ determinations based on the $w_2$, $w_3$
and $w_4$ FESRs, using the 3-loop, fixed-scale treatment of the $D=2$ 
OPE series favored by lattice data. The errors in the first half are
those associated with input uncertainties on the theory side, with 
``long corr'' labelling those associated with the sum rule/dispersive 
determinations of the small, doubly-chirally-suppressed continuum $us$ 
V and A channel $J=0$ subtractions. Combining these errors in quadrature 
yields a total theory error of $0.0004$ for each of the three cases. The 
errors listed in the second half of the table are those induced by the 
errors and covariances of the flavor $ud$ and $us$ V+A distributions. 
The experimental and total errors are both strongly dominated by the 
uncertainty on the $us$ V+A spectral integrals.

The excellent $s_0$-stability and agreement between the results 
from the different $w_N$ FESRs allows us to arrive at a final 
result for $\vert V_{us}\vert$ obtained by performing a combined 
fit to the $w_2$, $w_3$ and $w_4$ FESRs. We find
\begin{equation}
\vert V_{us}\vert = 0.2228(23)_{exp}(5)_{th}\ .
\label{adametzvusresult}\end{equation}
This is in excellent agreement with the results, $0.2235(4)_{exp}(9)_{th}$ and
$0.2231(4)_{exp}(7)_{th}$, obtained using the 2014 FlaviaNet experimental 
$K_{\ell 3}$ update~\cite{kell3andKratiosvus} and most recent
$n_f=2+1$~\cite{rbcukqcdfplus0} and $n_f=2+1+1$~\cite{fnalmilcfplus0} 
lattice results for $f_+(0)$. It is also compatible within errors with 
(i) the results, $0.2251(3)_{exp}(9)_{th}$ and $0.02250(3)_{exp}(7)_{th}$ 
obtained using the 2014 update for the experimental ratio
$\Gamma [K_{\mu 2}]/\Gamma [\pi_{\mu 2}]$~\cite{kell3andKratiosvus}
and the most recent $n_f=2+1$~\cite{rbcukqcdfkoverfpi} and 
$n_f=2+1+1$~\cite{fnalmilcfkoverfpi} lattice determinations of 
$f_K/f_\pi$ and (ii) the expectations of 3-family 
unitarity.{\footnote{An analogous analysis with $K\pi$ 
normalization not employing
the $B[K^-\pi^0\nu_\tau ]$ update of Ref.~\cite{adametzthesis}
yields $\vert V_{us}\vert = 0.2200(23)_{exp}(5)_{th}$. Of the $0.0024$
difference between this result and the conventional implementation
result~\cite{tauvusckm14} noted above, $0.0005$ results
from the use of $K_{\mu 2}$ for the $K$ pole contribution; the
remainder is due to presence of the $D=6,\, 8$ contributions 
not correctly accounted for by the assumptions of the conventional 
implementation. Note that the normalization of the two-mode
$K\pi$ sum produced by the $B[K^-\pi^0\nu_\tau ]$ update is
in good agreement with the results of the dispersive
study of $K\pi$ detailed in Ref.~\cite{aclp13}.}}

It is worth noting that, among the methods mentioned above, the one 
having the smallest theoretical error is, in fact, the FB FESR 
determination. This error is, moreover, as we have seen, a very 
conservative one. At present the experimental error on the FB FESR 
determination (resulting almost entirely from uncertainties in the 
$us$ exclusive mode distributions) is larger than those of the 
competing methods. We note, however, that the $us$ spectral
integral error is currently dominated by the uncertainty on the
branching fraction normalizations for the exclusive strange modes,
and hence systematically improvable through improvements in
these branching fraction values in the near future. In the longer
term, it will be important to complete the analysis of the exclusive
mode $us$ distributions not yet remeasured by the B-factory experiments
and to finalize the analyses of the covariances of those unit
normalized exclusive distributions which are not yet complete at present.

\acknowledgments{KM thanks the CSSM, University of Adelaide for
its hospitality during the time much of the work reported here
was completed.}

\end{multicols}

\vspace{-1mm}
\centerline{\rule{80mm}{0.1pt}}
\vspace{2mm}

\begin{multicols}{2}

\end{multicols}

\clearpage

\end{CJK*}
\end{document}